\documentclass[graybox,natbib,nosecnum]{svmult}
\bibpunct{(}{)}{;}{a}{}{,} 

\pdfoutput=1   
\usepackage{mathptmx}       
\usepackage{helvet}         
\usepackage{courier}        
\usepackage{type1cm}        

\usepackage{makeidx}         
\usepackage{graphicx}        
\usepackage{multicol}        
\usepackage{amsmath}	
\usepackage{amssymb}
\usepackage[bottom]{footmisc}
\usepackage[normalem]{ulem}	
\usepackage{hyperref}  

\usepackage{soul}

\makeindex            \usepackage{amssymb}

\newcommand{\vsini}{$V \sin i_\star$}   
 
\newcommand{\vmic}{$V_{\rm mic}$}
\newcommand{\vmac}{$V_{\rm mac}$}

\newcommand{\teff}{$T_{\rm eff}$}
\newcommand{\logg}{log\,$g_\star$}

\newcommand{\feh}{[Fe/H]}

\newcommand{\kms}{km\,s$^{-1}$}

\newcommand{\rstar}{$R_\star$}

\begin{document}

\title*{Methods to Characterize Exoplanet Host Stars from Spectroscopy}
\author{Carina M. Persson}
\institute{Carina M. Persson \at Chalmers University of Technology, department of Space, Earth, and Environment, Onsala space observatory, 439 92  ONSALA, Sweden, \email{carina.persson@chalmers.se}}
\maketitle

\abstract{A key to understand exoplanets is   characterization of their host stars. One of the most powerful tools to characterize  stellar properties like effective temperature, surface gravity and metallicity, is spectroscopy based on observations of stellar  atmospheres.   
This chapter   describes  the stellar parameters that can be derived from a spectrum   with examples of   well established  methods  and theoretical model atmospheres. 
Combined with photometry and parallax measurements, the outcome of the spectroscopic modelling can be used to derive 
 stellar radii and masses. }

\section{Introduction }
Exoplanets are intimately connected to their host stars through formation and evolution. 
In addition, detection and characterization of exoplanets   depend on detailed knowledge of their host stars since the current major detection techniques, transit photometry, and the radial velocity (RV) method detect planet sizes and masses relative to their host star as described in Radial Velocities as an Exoplanet Discovery Method (Jason T. Wright) and 
Transit Photometry as an Exoplanet Discovery Method (Hans Deeg and Roi Alonso). Uncertainties in a host star's parameters propagate directly to the planets.

Stellar modelling is normally based on two major techniques --  photometry   and spectroscopy. 
These methods are model dependent in contrast to the   direct measurements by  interferometry, eclipsing binaries, and asteroseismology.   

For eclipsing binaries \citep[e.g.,][]{1991A&ARv...3...91A,2010A&ARv..18...67T,2021A&ARv..29....4S}, 
and for the few large and nearby stars that enable  interferometric measurements \citep[e.g.,][]{2001ARA&A..39..353Q, 2023ARA&A..61..237E}, 
it is possible to derive the stellar radius with an accuracy of a few percent. 
The  timing of the duration of the eclipses of eclipsing binaries and their orbital velocities allows accurate estimates of their sizes. Similarly, 
stellar masses can be accurately determined for visual binaries  from observed separations from the common center of mass with Kepler's third law that relates their masses with observed orbital period and separation.
Asteroseismology can also be used to derive stellar mass, radius, and age  with a high precision  as described in detail in Characterizing Host Stars Using Asteroseismology by Mia Sloth Lundkvist, Daniel Huber,  V\'ictor Silva Aguirre, and William J. Chaplin. Recently, the seismic surface gravity of the star has also been used to obtain the effective temperature and metallicity, in particular with APOGEE \citep{2020ApJS..249....3A, 2022yCat.5154....0A}.

These  methods are, however, currently 
only possible to apply to a small subset of all stars. 
Spectroscopic measurements open a window to derive stellar parameters from a larger pool of stars than from   direct methods. 
High-resolution  spectroscopy  is a powerful tool that provides a wealth of information: the  effective temperature,  surface gravity,  chemical composition, and velocities. 
The stellar radius, and the luminosity, can then readily be derived from its spectral energy distribution (SED) via the spectroscopic parameters combined with photometry and parallax.

The spectroscopic parameters also serve as a base to model characteristics that normally cannot be directly measured, like mass and age, with stellar evolution models and the complementary tool isochrones 
\citep[e.g.,][]{2016ApJS..222....8D, 2018ApJ...856..125H}.  
An isochrone   is an evolutionary track  for a population of stars with different masses   on the Hertzsprung-Russell diagram  with the same ({\it iso}) age ({\it chrone}). The mass and radius can also be obtained from the spectroscopic parameters and empirical calibration equations \citep[e.g.,][]{2010A&ARv..18...67T,2010A&A...516A..33E, 2011MNRAS.417.2166S} albeit often with a higher uncertainty.  

The downside of high-resolution spectroscopy is that it is expensive; the observations are   time-consuming and also require  bright stars or a  large collecting area. 
Modelling of stellar parameters has therefore traditionally been performed by photometry because a much larger number of stars can  quickly be observed and analyzed. 
The effective temperature of a star can for instance be derived from color-color diagrams which have been scaled to stars measured with direct methods \citep[e.g.,][]{1989MNRAS.236..653B, 1996A&A...313..873A, 2004A&A...418..989N}. 
However, such models  have  in general higher uncertainties than based on spectroscopic measurements. In that respect they may only  be reliable in a statistical sense and not for individual stars. 
Thus for exoplanet host stars (for which direct measurements are not applicable), high-resolution spectroscopy is preferred.

This chapter begins with a short summary of a few basic requirements in order to obtain spectroscopic measurements of a star. It continues with an overview of the  parameters that can be extracted from a high-resolution spectrum and a few examples of  well-established methods. Some advantages and caveats are highlighted. The chapter ends with a brief summary of how to  combine spectroscopic parameters, photometric measurements, and  stellar evolution models to obtain stellar radii and masses. Modelling of stellar ages is described in Ages for Exoplanet Host Stars by Christensen-Dalsgaard and Aguirre.

\section{Instruments and spectral resolution}
Important properties of a spectrum are  high-resolution, high signal to noise (S/N), and wavelength coverage. 
There are several different types of spectrometers, and   many textbooks have been written    about this topic including advantages and challenges for different types of instruments    \citep[e.g.,][]{gray08}.  
The most successful type  in observations of exoplanets  is  echelle   spectrographs.
The main advantage   is the high resolution combined with a wide wavelength coverage obtained in a \mbox{single} exposure. 
Francesco Pepe describes the high-precision cross-dispersed echelle spectrographs for exoplanet research CORAVEL,
ELODIE, CORALIE, SOPHIE, and HARPS in High-Precision Spectrographs for Exoplanet
Research.

For the purpose of spectroscopic modelling of host stars, we want high resolution in order to resolve the  spectral lines.  
The    spectral resolution  $\Delta \lambda$ at wavelength   $\lambda$  is 
 \begin{equation} 
    \frac{\Delta  \lambda}{\lambda} = 
    \frac{1}{R}\ ,  
 \end{equation}
where      $R$ is the resolving power   of the spectrograph.
It can be translated into a velocity resolution  according to
 \begin{equation}
      \Delta V  = \frac{\Delta \lambda}{\lambda} c = \frac{c}{R}\ .  
 \end{equation}

Doppler broadening  of spectral lines due to thermal and turbulent motions of absorbing species in the atmospheres  
produces line widths of $\sim6$~\kms~for late-type stars.    
This corresponds to a spectral resolution of   50,000. 
Low-mass, slowly rotating stars can have line widths of only $\approx 1-2$~\kms, which  require $R \gtrsim 300,000$  to resolve the spectral lines and disentangle blended lines.

In terms of  high resolution, ultra-high precision,  and long-term  stability, 
the High-Accuracy Radial Velocity Planet Searcher \citep[HARPS;][]{Mayor2003HARPS} 
and  its decade younger sibling HARPS-North \citep[][]{Cosentino2012} 
 mounted on the 
ESO 3.6~m telescope (La Silla observatory, Chile)  and the 
Telescopio Nazionale Galileo (TNG) of Roque de los Muchachos Observatory 
(La Palma, Spain), respectively, have been  the leading instruments in detecting exoplanets over the last two decades. 
Both HARPS instruments  are fiber-fed cross-dispersed high-precision  echelle spectrographs  covering 380 to 690~nm. The spectroscopic resolution is approximately $115,000$ at visual wavelengths corresponding to a velocity resolution of 2.6~\kms. The design is  based on experience with the groundbreaking ELODIE and CORALIE instruments where the former was used to detect the first exoplanet 51\,Peg\,b \citep{MayorQueloz1995}.
Both HARPS spectrographs  can  be considered as the ``gold standard'' when searching for exoplanets in RV data and are also used for characterization of exoplanet host stars. 
When searching for exoplanets with the RV method, many    measurements are collected, sometimes over a period over many years. 
The individual spectra  can be   co-added  (after correcting for the periodic changes in   radial velocity) in order to increase the S/N enabling spectroscopic characterization of the host star.

In addition to the HARPS spectrographs, 
there are many other instruments used for exoplanet detection that can also  be used to characterize host stars e.g. ESPRESSO \citep{2021A&A...645A..96P}, FIES \citep{Telting2014}, HIRES \citep{Vogt94}, CHIRON \citep{Tokovinin2013CHIRON}, TULL \citep{Tull1995spectrograph}  with different resolutions operating at different wavelengths.

\section{Stellar Properties from Spectroscopy}
Only a few parameters characterize a
 stellar atmosphere: the effective temperature (\teff), the
surface gravity (\logg), the overall metal abundance ([M/H]), and atmospheric and rotational velocities.

The surface of a star is defined as the location where photons escape from the star which occurs at a characteristic optical depth of 2/3.  
This   occurs within the photosphere,  the innermost $\approx500$~km   of a star's atmosphere which overlies the opaque interior. 
The photospheric temperature and density varies with depth  and depend on the surface gravity as well as the abundances and opacity of the gases. 
The Sun's photosphere has a temperature that varies between 4400 and 6600~K  with an effective temperature of 5772~K,  while the  density is approximately \mbox{$3 \times 10^{-4}$~kg~m$^{-3}$},  
increasing with depth into the Sun. Other stars may have hotter or cooler photospheres. 
Spectral absorption lines   originate from different depths and opacities within the photosphere. Weak and optically thin spectral lines, and the line wings of optically thick lines, originate essentially in the same layer as the continuum. In contrast, the cores of  optically thick saturated lines  develop in higher layers. An example is  the core of the hydrogen~$\alpha$~line at 6562.81~\AA~which  originates in the hotter and less dense chromosphere which lies on top of the photosphere, where the assumption of local thermal equilibrium (LTE)  is no longer valid.

\subsubsection{Effective Temperature}

Stars are classified according to effective temperature  from the hot O-stars, also called early-type stars, to the cool M-stars (late-type stars). If a star is on the main sequence,  the effective temperature immediately signals which type of star it is along with typical mass and radius.  

Instead of choosing a particular depth to define a star's surface temperature, the effective temperature is defined  in terms of flux.  
The effective temperature is defined via the 
Stefan-Boltzmann  law    in terms of  the total power per unit area, radiated by the star    \citep[e.g.,][]{gray08}
\begin{equation} \label{Eq: Stefan-Boltzmann}
    \int_0^\infty  \mathcal{F}_\nu d \nu = \sigma T_\mathrm{eff}^4\ .  
\end{equation}
Here $\mathcal{F}_\nu$ is the total   flux passing through the star's surface and $\sigma$ is the Boltzmann constant. The effective temperature is thus the temperature of a black body having the same power output per unit area as a star. It is related to the   flux we measure at Earth, $f_\nu$,  via 
 \citep{gray08}
\begin{equation}  
    \int_0^\infty  f_\nu d \nu = \left ( \frac{R_\star}{d} \right )^2\int_0^\infty  \mathcal{F}_\nu d \nu = \left ( \frac{\theta_\star}{2} \right ) ^2 \sigma T_\mathrm{eff}^4\ ,  
\end{equation}
where the left integral is the   total radiative flux from the
star received at the top of the  Earth's atmosphere (bolometric flux), $R_\star$ is the stellar radius, $d$ is the distance, and $\theta_\star$ is the angular diameter of the star. 
The effective temperature can hence be derived by measuring the angular size of the star from interferometry and the received flux at Earth over a wide spectral range. 
By combining the angular size with distance  from parallax measurements, 
the linear radius of the star can be inferred for a wide range of spectral types nearly model independent except for the dependence on the adopted limb-darkening coefficients and bolometric correction.  
Alternatively, if the radius of the star is known from e.g.,  eclipsing binaries and the distance from parallax measurements, this will give the angular size   which then can be used to compute \teff. 
Since the target stars have to be
nearby  to measure their angular sizes,  interstellar absorption can be neglected. 
The infrared flux method  \citep[IRFM;][]{1977MNRAS.180..177B, 1980A&A....82..249B, 1990A&A...232..396B, 1989MNRAS.236..653B,  2006MNRAS.373...13C, 2010A&A...512A..54C} is based on observations of the angular size of the star and the measured infrared flux at the top of the Earth's atmosphere. 
The bolometric flux 
is   derived  taking into account the bolometric correction which gives  the effective temperature. 

In addition to the above methods, \teff~can also be derived from spectroscopy as described in the following section. However, 
it is worth highlighting that the  temperature derived from spectroscopy is a microscopic value which is close to, but   not exactly the same as the effective temperature due to its definition being a macroscopic description.

\subsubsection{Surface Gravity}
The surface gravity   is  an indication of the luminosity class of a star where V is the main sequence and I--IV is different types of giants. Thus surface gravity contains information of the size and age of a star. A low surface gravity immediately indicates that the star has left the main sequence.

The surface gravity of a star is defined by \citep[e.g.,][]{gray08}
\begin{equation}
    g_\star = g_\odot \frac{M_\star}{R_\star^2}\ , 
\end{equation}
where $g_\odot$ is the surface gravity of the Sun ($2.740\times 10^4$~cm~s$^{-2}$) and the  radius, $R_\star$, and mass, $M_\star$, of the star are in solar units.
The surface gravity is commonly measured in a logarithmic scale, \logg, where the solar value is 4.44.

The surface gravity determines the   gas density in the photosphere and is the spectroscopic parameter that has the highest impact on the stellar radius.  Unfortunately, \logg~is  often poorly constrained by spectral analysis. Since 
the uncertainties of \teff~and metallicity can be strongly correlated with  surface gravity this can lead to a significant source of systematic error in some analysis techniques.

\subsubsection{Metallicity}
A stars chemical composition is an outcome of the  nucleosynthesis by previous generations of stars. This is important when reconstructing   star   and planet formation   history in our Galaxy. 
There is a large   variation of metallicity, i.e., the abundance of all elements heavier than helium denoted with [M/H], in the Milky Way up to about twice the solar value to hundreds of thousands of times lower than the solar value 
\citep[e.g.,][]{Christlieb2004ChemicallyAncientStar,Haining2022MetalPoorStars,Nepal2024MetalRichStars}.
 
The   chemical composition of a star  is commonly fixed 
to the overall metallicity of a star relative to the Sun. The \citet{2007SSRv..130..105G},  \citet{Asplund2009} and \citet{2003ApJ...591.1220L} abundance scales for the Sun are currently the most   adopted. 
However,   individual abundances of a star may not follow solar composition   
and may   require  modelling of individual elemental abundances. 
Abundances are generally  measured on a logarithmic scale normalized to the Sun  where zero equals the Sun’s metallicity. In the case of iron we have   
[Fe/H]$_\star$   = log(Fe/H)$_\star$  -  log(Fe/H)$_\odot$. For example, an iron abundance of [Fe/H]$\, = -0.5$ means that the abundance is  $10^{-0.5}$ relative to the Sun. 
Since iron is by far the most abundant species in a stellar atmosphere after hydrogen and helium, measurements of   iron   have become a proxy for the   metallicity. 

Stellar abundances are also important when modelling exoplanet interiors in particular rocky super-Earths without significant gaseous envelopes. The degeneracy of interior composition inferred from radius and mass measurements can for this type of planet be reduced assuming an interior structure with a differentiated iron core and a rocky mantle.  
In these cases, the host star abundances are often used as a proxy of the primary planet-building elements Fe, Mg, and Si, which are expected to be reflected in the planet composition, planet interior, and core mass fraction \citep[][]{Dorn15,2023A&A...677A..14A}.

\subsubsection{Velocities}

 Thermal widths of spectral lines are only a fraction of the observed line widths for  dwarf stars hotter than spectral type K0. The line widths    are instead mainly  governed by 
Doppler shifts    produced by  motions of the star's   photospheric gases. The radial velocity of the star is only shifting the wavelengths of all spectral lines in the observed spectrum compared to the observations and can easily be corrected for.  The  velocity that   dominates the  line shape and width for hot stars is the 
projected equatorial   rotational velocity of the star, \vsini, where $i_\star$ is the inclination of the stellar rotation axis
relative to the line of sight. It can be measured via the full width at half maximum (FWHM) of a large number of optically thin and unblended lines not sensitive to pressure broadening.  The line shapes are, however,    also   affected by turbulence from 
  convective motion, granulation, high-order pulsations, stellar activity, and other types of local flows in the photosphere. 
  Turbulence   is represented in the models  by  the macro-turbulent velocity (\vmac) that  describes     
  motions on scales larger than the mean free path within the photosphere that induce a change in the line shape; 
  and  the micro-turbulent velocity    (\vmic) that describes motions
on scales smaller than the mean free path leading to increased line opacity \citep{gray08, Bruntt2010b, Doyle2014}. 
  The latter velocity is a ``fudge'' factor 
  originally introduced to reconcile observed and predicted equivalent widths \citep{1978stat.book.....M}. 
  It includes all remaining types of broadening mechanisms and is at present standard to include in  analyses of solar-type stars. 
    Both turbulent velocities depend on   temperature and to a lesser extent   on surface gravity. The micro-turbulent velocity   has a width of the order of 1~\kms~for dwarfs and several~\kms~for giants. 
  For low-mass stars, \vmac~and \vsini~have comparable widths of the order of a few \kms. As a reference, the Sun's \vsini~is 2~\kms~at the equator while hotter stars have much higher rotational velocities (tens to hundreds of \kms).

\section{Methods for Spectroscopic Modelling} \label{Section: Methods for spectroscopic modelling}

There are several ways to model a spectrum which can be divided into two main groups. The first is based on spectral synthesis. Here 
observations are fitted to a synthetic spectrum of stellar atmosphere models by comparison of line profiles. The second is   a line-by-line analysis based on measured strengths of observed spectral lines  and their equivalent widths  (EWs).  Detailed description of the physics can be found in many textbooks e.g., \citet{gray08}.
Spectroscopic observations can also be compared to a library of spectra of well-characterized stars via for example interferometric measurements or spectroscopic binaries. 
 
One major problem in spectroscopic analysis is to accurately determine the continuum which can introduce large errors. This is particularly difficult  for poor spectra with low spectral resolution or low S/N. It also depends on the spectral type of the star and the wavelength region. The number of spectral lines   increases toward shorter wavelengths for all types of stars. 
In addition, late-type stars have    a  much higher density of spectral lines  than early-type stars arising from both atoms and molecules leading to blending and confusion of  the continuum location.  The higher temperatures of early-type stars  ionize a large fraction of their atoms, leading to significantly fewer spectral features than low-mass stars. Differences in rotational velocities also affect the spectral line density. 
In contrast to late-type stars, 
the early types have  very  high rotational velocities   which leads  to very broad   spectral lines that smear out spectral features. 
Thus solar-type stars (FGK) are the easiest stars to model, while high- and low-mass stars often entail a significantly higher degree of difficulty in the modelling.  
M-dwarfs have in addition generally a much longer period of high  stellar activity than FGK stars,   exacerbating the problems \citep[e.g.,][]{2023A&A...675A.168M}.  
This is unfortunate since M-dwarfs are popular exoplanet host stars due to their small masses and sizes which increase the exoplanet signals.

Care must   be taken when selecting which spectral lines to model. 
A large set of narrow, non-blended spectral lines   are preferred (unless modelling pressure broadened line wings, see below). 
If a spectral line becomes optically thick, the abundance of a species stops growing linearly with absorption depth. The characteristics of an optically thick line is a saturated line center which flattens the bottom and broaden the line wings. 
Not all optically thin spectral lines may, however, be useful since a large number comes with poorly determined atomic parameters which are needed to compute synthetic spectra. This can be circumvented for solar-type stars    if adopting new atomic parameters after comparing the lines from observations of the Sun.

\subsection{Fitting Observations to Synthetic Spectra}
Computations of a synthetic spectrum  requires a model atmosphere based on solutions to the stellar structure equations to synthesize a spectrum. Most stellar atmosphere models are  pre-calculated and tabulated on grids  describing the profiles of the temperature, surface gravity, and abundances  as functions of atmospheric depth. 
Each layer in the model atmosphere   is contributing to the formation of absorption line profiles in the final spectrum. 
Some widely used atmospheric model atmospheres are   Atlas12 \citep{Kurucz2013}, Atlas9  \citep{kur93cd13,2002A&A...392..619H},   MARCS \citep{Gustafsson08} for cool and giant stars, and   LL models \citep{2004A&A...428..993S} for hot main sequence stars. Line lists  of atomic and molecular  data needed in the computation can   be provided by the Vienna Atomic Line Database
  \citep[VALD3;][]{pisk95, Ryabchikova2015}. 
  
A radiative transfer code then computes   a synthetic model spectrum of the star for the chosen set of stellar parameters (\teff, \logg, \feh, \vsini, \vmic, \vmac) which are  matched against  the  observed spectra   based on 
the spectral line shapes and strengths.  
The parameters generally affect either the line strength (\teff, \logg, abundances, and \vmic) or the line shape (\vmac, \vsini, and the instrumental resolution). Degeneracies are stronger within the subsets. 
 
 The dependence of the line profile on the \vmac~parameter is  broadened line wings and a cusp-shaped core. 
 Unfortunately, disentangling the effect on the line profile from \vsini~and \vmac~is difficult, leading to a degeneracy between the two. Prior information of  
\vsini~may be obtained from time-resolved photometry,    available  for the known transiting planets,   or from 
 asteroseismology \citep[cf.][]{Doyle2014}.   
If no  prior information   is available, 
 calibration equations of both turbulent velocities are often used \citep[e.g.,][]{Bruntt2010b, Doyle2014} which allows \mbox{modelling} of \vsini.  
  Note that in order to properly model the different velocities above, it is  imperative to   take into account how the spectrograph itself broadens the lines. Spectral lines   can also be  pressure broadened through various mechanisms which further broaden the lines,   e.g., the Balmer lines. Details of the various broadening mechanisms can be found in many  textbooks  such as  \citet{gray08}.

There are many software that computes synthetic spectra to be used as a model constraint to interpret the observed spectrum. 
 The popular open-source spectroscopic tool   iSpec     
 \citep{ispec2014,ispec2019}     supports several of the most
well-known radiative transfer codes such as  Spectroscopy Made Easy \citep[{\tt SME}; ][]{vp96, pv2017, 2023A&A...671A.171W},  {\tt SPECTRUM} \citep{Gray1994},  {\tt Turbospectrum} \citep{2012ascl.soft05004P, 2012A&A...544A.126D, 2023A&A...669A..43G},  {\tt Synthe/WIDTH9}  \citep{kurucz93_synthe}, and  {\tt MOOG} \citep{MOOGphd} where the latter is based on the equivalent width method.  
The codes often adopt LTE and a plane-parallel geometry as default  motivated by the fact that
for main sequence stars, the photosphere constitutes  $\ll1$~\% of the stellar radius.
A spherically symmetric geometry is on the other hand required for giant stars where the atmosphere makes up a substantial portion of its radius \citep{2006A&A...452.1039H}. 
 However, since complex interaction of  gas particles and the nonlocal radiation fields   leads to deviations from LTE in the atmospheres of FGKM-type stars, some of the softwares have also the option of nonlocal thermodynamic equilibrium (NLTE). For instance,    {\tt SME} includes NLTE departure coefficients for the MARCS and LL models atmospheres \citep{pv2017}, and  {\tt Turbospectrum} also have the option of NLTE \citep{2023A&A...669A..43G}.

Fitting can in several of the software be made for one or several parameters at the same time using a $\chi^2$-minimization algorithm. However, care must be taken when solving for several parameters simultaneously \citep{2012ApJ...757..161T} due to degeneracies and difficulties in the modelling. 
Different initial assumptions of one free parameter at a time can therefore be made in the fitting process to iterate to the final solution, thereby mitigating  degeneracies.  
Also, since the surface gravity is often difficult to constrain,  it is thus advantageous to have  external information about the stellar density that can facilitate modelling. 

\begin{figure}
\centering
\includegraphics[scale = 0.4]{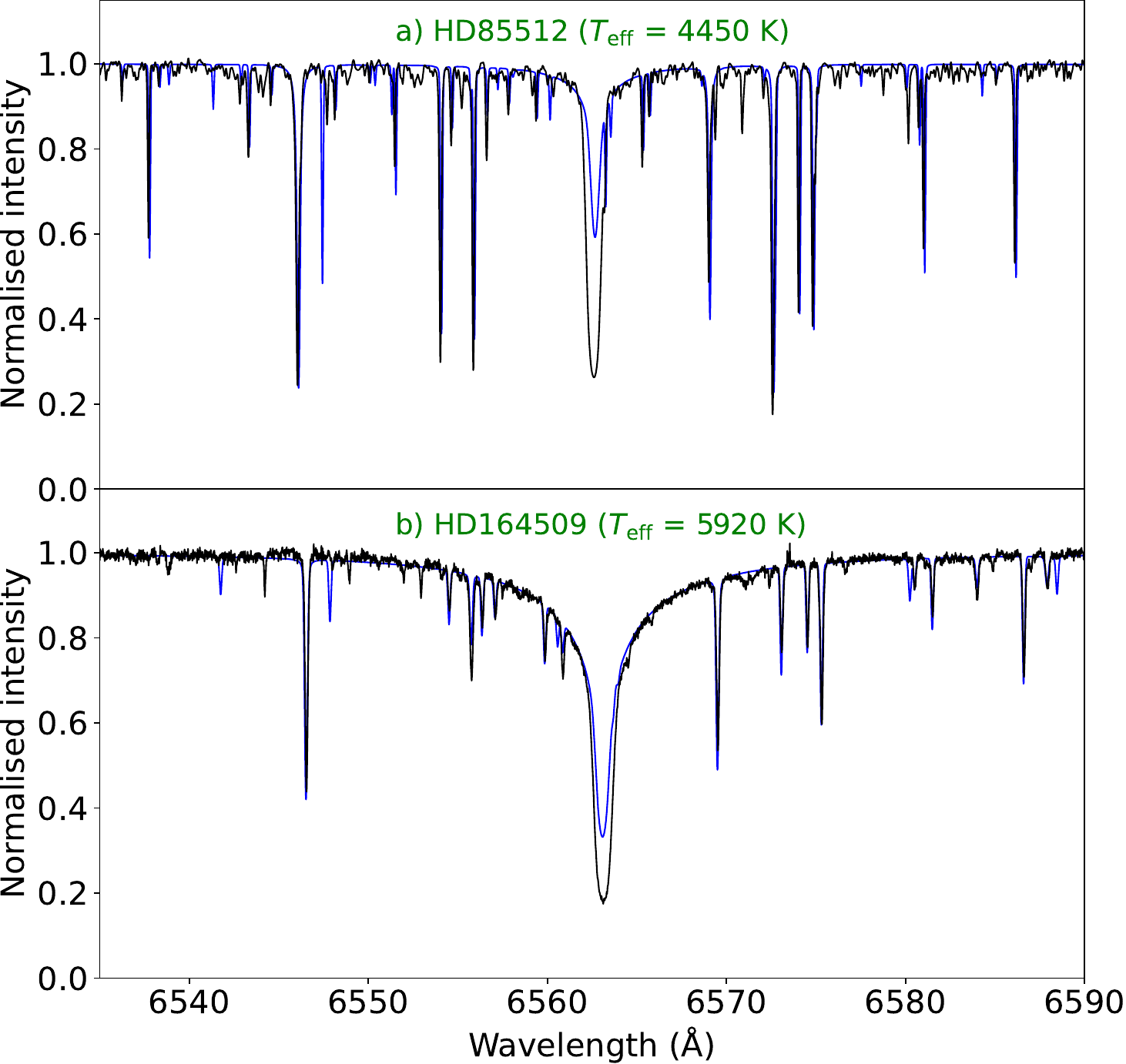}
\caption{Examples of {\tt SME} modelling of H$\alpha$ using the MARCS (LTE) stellar atmosphere model  toward a K5V (HD85512) and a G0V (HD164509) host star observed with 
HARPS.   The observations are plotted in black and the models in blue. 
The  line \emph{wings} of H$\alpha$ are very sensitive to the effective temperature. Toward the K5V star they have almost disappeared, while they are very broad toward the G0V star.    
}
\label{fig: Teff Halpha}   
\end{figure}

\begin{figure}
\centering
\includegraphics[scale = 0.4]{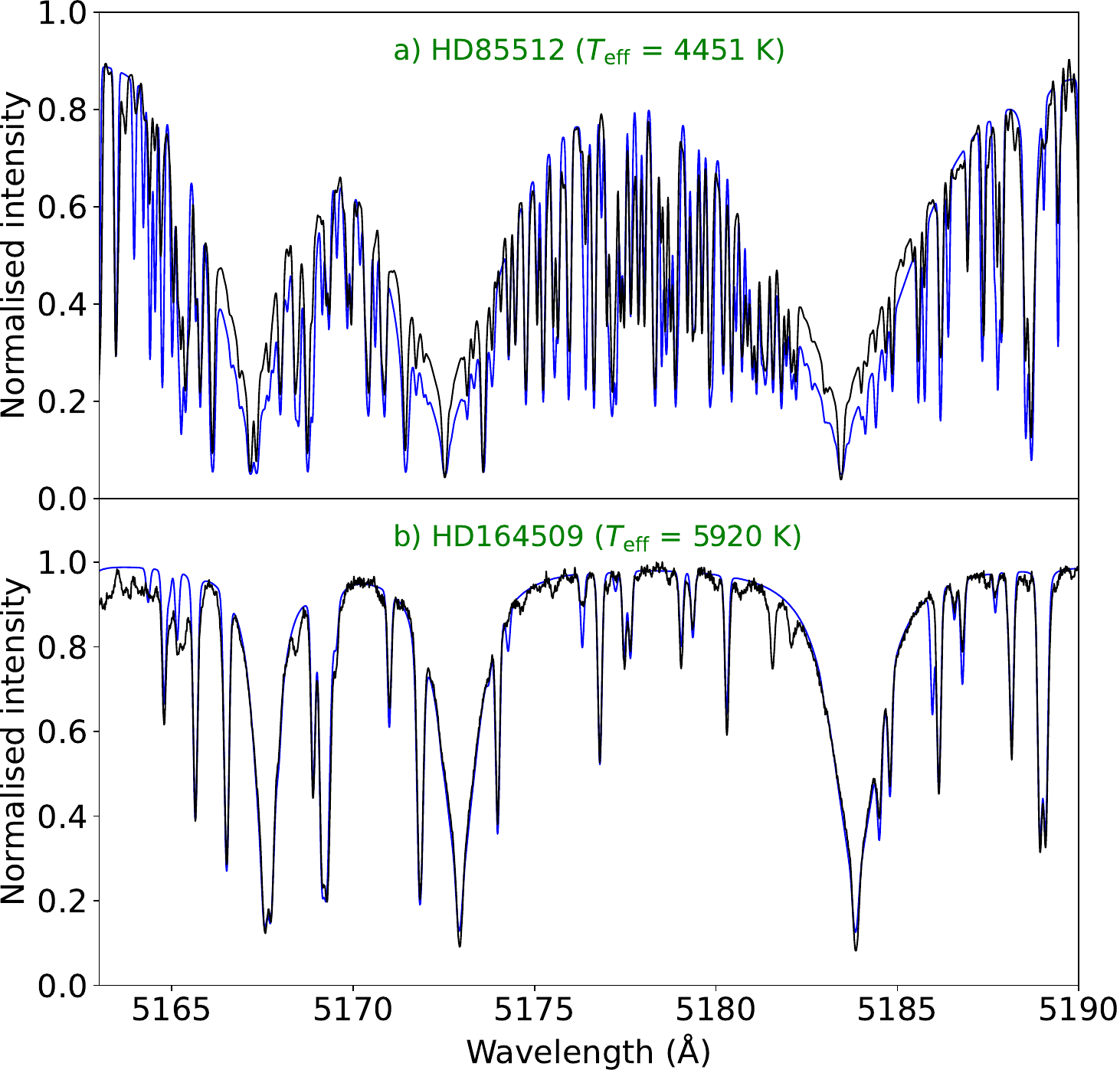}
\caption{
Examples of {\tt SME} modelling of 
the Mg~I triplet (5167.33~\AA, 5172.70~\AA, and 5183.62~\AA) toward  
the same stars as in Fig.~\ref{fig: Teff Halpha}. The  line \emph{wings} are sensitive to \logg. 
Note the difference in spectral line density and line  blanketing in HD85512 which introduce severe problems in the spectral modelling. 
}
\label{fig: logg Mg}        
\end{figure}

Spectroscopic modelling can take advantage of the sensitivity of certain spectral lines to specific parameters. 
For instance, all Balmer lines exhibit  pressure (collisionally) broadened line wings,  which makes their 
profiles   strongly sensitive to temperature  \citep[e.g.,][]{fuhrmann93, fuhrmann94, Barklem2000b_hydrogen-self-broadening, Barklem2002} as shown in Fig.~\ref{fig: Teff Halpha}.   
The Balmer line  wings are   formed in the deepest photospheric layers likely close to the 
LTE.  
The metal-line blending  increases from H$\alpha$ at 6562.81~\AA~to H$\delta$ at 4101.75~\AA, which perturbs the higher transitions of the Balmer line profiles making H$\alpha$  the best choice. 
For late F, G, and early K-stars, H$\alpha$   is   very  insensitive to \logg~and abundance making its line wings an excellent \teff~indicator.  
Two examples of H$\alpha$ line profiles toward a K5V   and a G0V host star are shown in Fig.~\ref{fig: Teff Halpha}. It is obvious that the line wings have almost completely disappeared in the K5V  spectrum, while prominent and very broad toward   the G-star. 
For M-dwarfs where the line wings of H$\alpha$ are absent,   TiO lines  can also be used as a \teff~indicator \citep{1998ApJ...498..851V}. 

For  late F- and G-dwarfs 
there are several   pressure-sensitive lines that can be used to constrain \logg: The Mg~I triplet at 5167.33~\AA, 5172.70~\AA, and 5183.62~\AA~\citep[e.g.,][]{fuhrmann97b,ValentiFischer2005} as shown in  Fig.~\ref{fig: logg Mg}, and the Ca~I lines at 6122.23~\AA~and 6162.18~\AA~ \citep[e.g.][]{gray08}. For lower gravity  or higher temperature, the line wings disappear due to lower photospheric density or ionization. 
The  Mg~I lines  are, however, very wide which makes normalization difficult. Two of the lines lie very close without 
 continuum between them, and there are   numerous overlying narrow metal lines in particular for late-type stars. In these cases, only the 5183.62~\AA~line can be used reliably. 
 Figure~\ref{fig: logg Mg} shows the Mg~I triplet line profiles   toward the same host stars as in Fig.~\ref{fig: Teff Halpha}. 
 As already seen in Fig.~\ref{fig: Teff Halpha}, 
 the spectral line density is much higher toward the K5V star   than the G0V star. However, the effect is even more pronounciated at shorter wavelengths. 
In addition to surface gravity, the Ca and Mg pressure broadened line wings are, however,   to some degree also sensitive  to temperature and    abundance. The procedure is therefore to model the temperature first, e.g., via  H$\alpha$, and the abundance       via    narrow, unblended  Ca (e.g., 6156.02~\AA,  6166.439~\AA, 6169.042~\AA, 6455.985~\AA) and Mg (e.g., 5711.09~\AA)  lines    and iterate to a solution.

The broad line wings of Na~I D-lines (5889.97~\AA~and 5895.94~\AA) are sensitive to both \teff, \logg~(and the Na abundance). Hence these lines can be used to check the final model   for consistency.

More details and examples of  spectral modelling   can for instance be found in  \citet{ValentiFischer2005},   \cite{2014A&A...564A.133J}, and \citet{Brewer2016}.

\subsection{Fitting Equivalent Widths}
In contrast to synthetic spectral synthesis methods, the  equivalent width   (EW) method begins   with the observed spectrum by measuring
the strengths  of selected  absorption lines which are translated into
individual line abundances.
The method is based on theoretical atmosphere models and  excitation and ionization equilibrium which determines the 
 population in a certain level of an atom or an ion.  

The EW of an absorption line is a convenient measurement of its strength. 
It  is defined as the width of a rectangle  reaching up to the continuum (with length one) having  the same  area as the spectral line.
 Measuring  EWs of well-defined weak neutral iron  (Fe~I)   and ionized iron (Fe~II) lines is a traditional method to model stellar spectroscopic parameters since there are numerous  iron lines   in stellar \mbox{spectra,} many with accurate atomic parameters.  
Metal lines can   be very sensitive to temperature although  large variations between lines exist. Depending on spectral type, neutral metals are often used as a temperature indicator in solar-type stars, while  ionized metal lines are better  tracers in early-type stars.
 
Assuming LTE, the  
ratio of population in two levels $n$ and  $m$ of an atom (or ion)   of a   species   can be computed with the Boltzmann equation \citep[e.g.,][]{gray08, 2017imas.book.....C}  
\begin{equation}
    \frac{N_n}{N_m} = \frac{g_n}{g_m} e^{-(\chi_n - \chi_m)/kT}\ ,  
\end{equation}
where 
$g_n$ and $g_m$ are the statistical weights of the two levels,    $\chi_n$ and $\chi_m$  are the  corresponding  excitation potentials, $k$ is the Boltzmann constant, and $T$ is the temperature. Different line ratios have different sensitivity to temperature.

As the temperature increases so will ionization which occurs quite abruptly once the threshold temperature is reached. In a star's photosphere, elements exist mainly in just two ionization stages, which can be computed with the Saha equation \citep[e.g.,][]{gray08, 2017imas.book.....C}    between  ionization states $i$ and $i+1$ as 
\begin{equation}
    \frac{N_{i+1}}{N_{i}} P_e = \frac{(2\pi m_e)^{3/2} (kT)^{5/2}}{h^3} \frac{2u_{i+1}(T)}{u_{i} (T)} e^{-I/kT}\ , 
\end{equation}
where the electron pressure is $P_e = N_e kT$ (indicator of the surface gravity), $m_e$ is the electron mass, $h$ is Planck's  constant, $u (T) = \sum g_i\, e^{-\chi_i/kT}$ is the partition function, and  $I$ is the ionization energy.

The EW method requires measurements of a large number of equivalent widths either  measured by direct integration over the entire  line  or by fitting a Gaussian profile (or a   Lorentzian   profile that may fit optically thick lines better). 
Measuring EWs manually is, however, very time-consuming and therefore several software  have been developed to automate the measurements for example  
ARES  
\citep{sousa07} and DAOSPEC \citep{2008PASP..120.1332S}. 

Weak and optically thin iron lines   depend mainly on \teff~and the iron abundance and less on \logg~and \vmic. 
In order to avoid abundance dependence with the EW method, it is best to choose two lines from the same element. However, since continuum normalization is often a large source of error it may sometimes be necessary to use pairs of lines at nearby wavelengths. In these cases, pairs of similar elements like Fe, V, and Ti that normally have similar abundances, are often used. Different sets of lines are chosen for different spectral types since they are useful in different temperature ranges.

The effective temperature is constrained in the following modelling by the 
 correlation between the excitation potential and the iron abundance of each individual line, the microturbulence is constrained by the correlation between the     abundance of each line  and the reduced equivalent width, while the surface gravity is constrained by   the ionization balance. The parameters are adjusted until there are no correlations left and all individual abundances are the same \citep{2014dapb.book..297S}.

An example of a widely used   open 
radiative transfer code   based on the EW method is MOOG  
\citep{MOOGphd}. This software 
uses   a grid of the ATLAS9 \citep{kur93cd13} plane-parallel model atmospheres to  produce model spectra which is compared to the measured   EWs of individual Fe~I and Fe~II lines. 
 The equilibrium conditions  are solved simultaneously to derive   \teff, \logg, [Fe/H], and \vmic. A good description of the process is found in \citet{2014dapb.book..297S}. It has been used by for instance \citet{2021A&A...656A..53S} together with ARES for a 
homogeneous spectroscopic characterization of  almost one thousand  exoplanet host stars (the SWEET-Cat online catalogue). 
Another example of a software based on the EW method is ODUSSEAS \citep[Observing Dwarfs Using Stellar Spectroscopic Energy-Absorption Shapes;][]{ODUSSEAS2020}. This code uses the machine learning Python package scikit-learn to offer a
quick and automatic derivation  of \teff~and [Fe/H] for M dwarfs  from optical spectroscopy.

When comparing the outcome of the EW and  models based on synthetic spectra, 
\teff~and \logg~normally agree within the uncertainties and also with results
from 
asteroseismology  within 100~K and 0.1~dex, respectively. The methods, however, do not perform equally well for all spectral types. Since the EW method is based on differential analysis with respect to the Sun, it can be applied to FGK stars with \teff\,$\approx 4500-6400$~K  \citep[][]{2011A&A...533A.141S}. Modelling based on synthetic spectra and the line shapes is also sensitive to spectral type since
some of the fundamental traits, e.g., the broad H$\alpha$ line wings, can only be made for solar-type stars since the line wings disappear for colder and hotter stars. In these cases it may still, however, be possible to use other spectral lines (e.g. TiO for low-mass stars).
Furthermore, 
the EW method cannot be used   on
fast-rotator stars  because of the severe line blending. For these stars, it may still be possible to use the synthetic method. 
Another difference is that the EW method is normally much faster than the synthetic method, while synthetic models provide a more complete description of the star.

\subsection{Empirical Methods}
Instead of the above methods, it is also possible to compare an observed high-resolution spectrum to a spectra of well-characterized  stars and in this way derive stellar parameters.  
An example of such software is Spechmatch-emp \citep{2017ApJ...836...77Y}   that     compares  an observed optical spectrum   with a dense empirical spectral library thereby deriving \rstar~(or \logg), \teff, and \feh. The library contains 404 stars observed with the HIRES instrument with $R \approx 60,000$  at the Keck telescope.
The properties of all the library stars have been derived from asteroseismology, interferometry, spectrophotometry, and LTE spectral synthesis and represent spectral types from approximately M5 to F1. This method performs very well  for the difficult late type stars (K4 stars and later) which are challenging for other methods reaching accuracies of 10~\% in stellar radius, 70 K in effective temperature, and   0.09 dex in metallicity ([Fe/H]).  The software and the library are publicly available.

\section{Stellar Radius and Mass}
Once we have obtained the stellar spectroscopic parameters for our host star we now want to find out the radius and the mass of the star.

\subsection{Radius from Spectroscopy and   Spectral Energy Distribution}  
\begin{figure}
\centering
\includegraphics[scale = 0.38]{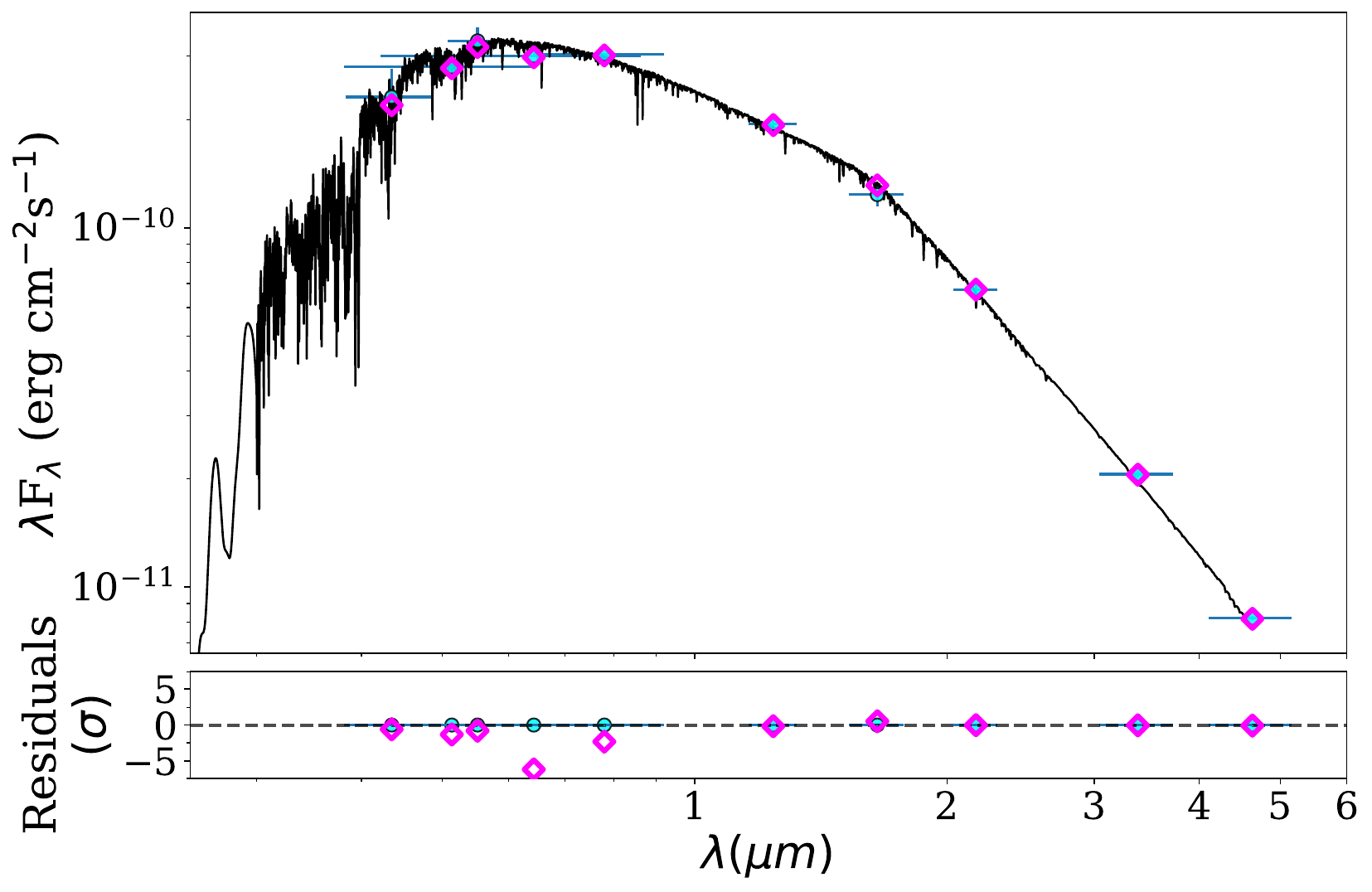}
\caption{Example of a spectral energy distribution (SED) fit  of the host star TOI-2196  discovered by the Transiting Exoplanet Survey Satellite \citep{Carina22}.
The observed photometric measurements are plotted with   blue circles and the   effective widths of the passband are represented with horizontal bars. The   {\tt {Phoenix}} \citep{2013A&A...553A...6H} atmosphere model is outlined in black and the magenta diamond  symbols are the corresponding model fluxes. 
}
\label{fig: SED}        
\end{figure}

When   spectroscopic parameters and photometric measurements are available  it is  straightforward to obtain the stellar radius from a fit of the observed magnitudes in different bands to its  spectral energy distribution   (SED). In addition to  magnitudes, the SED also depends on the distance (most accurately computed from the observed parallax), the spectroscopic parameters (derived from spectroscopic modelling), the extinction along the line of sight, and the stellar radius which is a free parameter in the fit.
Parallax measurements have been performed by the European space missions Hipparchos  
\citep{1997SSRv...81..201V} and    Gaia  
\citep{2016A&A...595A...1G} launched in 1989 and 2013, respectively. The latter mission   has  provided the community with  astrometric and photometric measurements of almost two billion stars in the Milky Way. Gaia has also delivered excellent measurements of the magnitudes in the Gaia optical band. Observations   will end in early 2025.    
Figure~\ref{fig: SED} shows an example of a SED fit with the {\tt {Phoenix}} \citep{2013A&A...553A...6H} atmospheric model grid. An example of a publicly available software that automatically 
fits broadband photometry to six different stellar atmosphere models  using Nested Sampling algorithms
is the {\tt {ARIADNE}}  
\citep[][]{2022MNRAS.513.2719V} software.

\subsection{Stellar Mass}
Unless the star is a binary, the mass cannot usually be determined directly but can be modelled via stellar evolution models.  Some of the most popular   models are  BaSTI \citep{2018ApJ...856..125H}, Padova    \citep{Bertelli08Padova,Bertelli09Padova}, DESP  \citep[the Dartmouth Stellar Evolution Database;][]{Dotter2008}, MIST \citep[MESA Isochrones and Stellar Tracks;][]{2016ApJ...823..102C},  
PARSEC \citep[the PAdova and TRieste Stellar Evolution Code;][]{Bressan2012PARSECisochrones}, 
and   Y$^2$  \citep[Yonsei-Yale;][]{Yi2003YonseiYale,Demarque2004YonseiYale}.  A quick way to obtain a Bayesian estimation of both mass and radius based on \mbox{PARSEC}  and MESA \citep{2017MNRAS.467.1433R}  models are the web interfaces {\tt Param1.3} and {\tt Param1.5} 
\citep{daSilva2006}. Another publicly available software 
that provides a simple interface for MCMC fitting of MIST stellar  model grids 
is {\tt isochrones}  
\citep{2015ascl.soft03010M}.

A complementing method is to use a set of  stars with well-known masses and radii, e.g. through interferometry and eclipsing binaries, to derive empirical calibration equations. Such equations can give the mass and radius of a star given a set of stellar atmosphere values \citep[e.g.,][]{2010A&ARv..18...67T,2010A&A...516A..33E, 2011MNRAS.417.2166S}.

\subsection{Final Checks}
If a planet is transiting, a final important  check can be made. The  stellar density  obtained from the above     radius and mass  
 can be checked against  the  density  obtained from transit photometry and Kepler's third law. Assuming a circular orbit we have \citep{seager03} 
     \begin{equation} \label{Eq: stellar density from transits}
\rho_\star = \frac{3\pi}{GP_\mathrm{rot}^2} \left (\frac{a}{R_\star}  \right )^3\ ,  
     \end{equation}
where $G$ is the gravitational constant, $P_\mathrm{rot}$ is the orbital period, $a$ is the semi-major axis,  and $R_\star$ is the stellar radius. If this density does not agree with the above derived value, you need to  check your modelling again. 
Equation~\ref{Eq: stellar density from transits} can be modified to include the eccentricity if known \citep[e.g.,][]{2011ApJ...726..112T}. 

Another checkpoint is to compare  the stellar mass computed from the spectroscopic \logg~combined with \rstar~which should be consistent with the final mass from other methods.
 
A final remark: If possible, it is best to use several methods  to make sure that the derived stellar parameters  are  robust and  consistent.

\section{Cross-References}
\begin{itemize}
\item Ages for Exoplanet Host Stars (J\o rgen Christensen-Dalsgaard and V\'ictor Aguirre B\o rsen-Koch).
 \item Characterizing Host Stars Using Asteroseismology (Mia Sloth Lundkvist, Daniel Huber,  V\'ictor Silva Aguirre, and William J. Chaplin). 
    \item High-Precision Spectrographs for Exoplanet Research:
CORAVEL, ELODIE, CORALIE, SOPHIE and HARPS (Francesco Pepe).
 \item Radial Velocities as an Exoplanet Discovery Method (Jason T. Wright).
   \item Transit Photometry as an Exoplanet Discovery Method (Hans Deeg and Roi Alonso).
\end{itemize}

\bibliographystyle{spbasicHBexo}   
\bibliography{references}  

 \appendix

\end{document}